\documentclass[aps,twocolumn,floats,floatfix,amssymb,nofootinbib,prd,superscriptaddress]{revtex4-1}
\usepackage{graphicx,amssymb,amsmath,amsthm,amsfonts,epsfig}
\usepackage[linktocpage,colorlinks=true,citecolor=OliveGreen,linkcolor=Maroon,urlcolor=Maroon]{hyperref}
\usepackage{appendix}
\usepackage{tensor}
\usepackage[normalem]{ulem}
\usepackage[dvipsnames, usenames]{xcolor}
\usepackage{nicematrix}
\usepackage{xr-hyper}
\definecolor{linkcolor}{rgb}{0.0,0.3,0.5}
\usepackage[all]{hypcap}
\usepackage[T1]{fontenc}
\usepackage[utf8]{inputenc}
\usepackage[usenames,dvipsnames]{xcolor}

\definecolor{romared}{RGB}{142,0,28}

\newcommand{\be}{\begin{equation}}
\newcommand{\ee}{\end{equation}}

\def\be{\begin{equation}}
\def\ee{\end{equation}}
\newcommand{\beq}{\begin{eqnarray}}
\newcommand{\eeq}{\end{eqnarray}}

\usepackage{makecell}
\usepackage{soul}

\newcolumntype{Y}{>{\centering\arraybackslash}X}

\definecolor{cornellGreen}{HTML}{6EB43F}

\definecolor{cornellGreen}{HTML}{6EB43F}
\definecolor{cornellRed}{HTML}{B31B1B}

\begin{document}

\title{Total absorption of tailored incoming signals by black holes}

%
\author{Furkan Tuncer}
\affiliation{Center of Gravity, Niels Bohr Institute, Blegdamsvej 17, 2100 Copenhagen, Denmark}
\affiliation{Bilkent University, Dept. of Physics, 06800 Bilkent, Ankara, Turkey}

\author{Vitor Cardoso}
\affiliation{Center of Gravity, Niels Bohr Institute, Blegdamsvej 17, 2100 Copenhagen, Denmark}
\affiliation{CENTRA, Departamento de F\'{\i}sica, Instituto Superior T\'ecnico -- IST, Universidade de Lisboa -- UL,
Avenida Rovisco Pais 1, 1049 Lisboa, Portugal}

\author{Rodrigo Panosso Macedo}
\affiliation{Center of Gravity, Niels Bohr Institute, Blegdamsvej 17, 2100 Copenhagen, Denmark}

\author{Thomas F.M.~Spieksma}
\affiliation{Center of Gravity, Niels Bohr Institute, Blegdamsvej 17, 2100 Copenhagen, Denmark}
\begin{abstract}
We uncover a new class of phenomena in gravitational physics, whereby resonances in the complex plane can be excited via tailored time-dependent scattering. We show that specific forms of temporal modulation of an incoming signal can lead to complete absorption for the entire duration of the scattering process. This, then, makes stars and black holes truly black. Such ``virtual absorption'' stores energy with high efficiency, releasing it once the process finishes via relaxation into the characteristic virtual absorption modes -- also known as total transmission modes -- of the object. While such modes are challenging to obtain and four-dimensional black holes have a restricted set of solutions, we also show that higher dimensional black holes have a complex and interesting structure of virtual absorption modes.
\end{abstract}
\maketitle
\section{Introduction}
Scattering experiments have long been central to physics, providing a powerful means of probing the internal structure and properties of physical systems. Traditionally, a scattering experiment such as Rutherford's study of atomic nuclei, is analysed within a stationary-wave approach, where a monochromatic plane wave -- decomposed into its different multipolar components -- interacts with a target and is partially reflected, transmitted, or absorbed~\cite{Rutherford:1911zz,Gordon:1928,Schiff:1968}. When the wave frequency matches a bound or quasi-bound state of the scatterer, a \textit{scattering resonance} occurs, leading to sharply enhanced absorption. A classic example is total absorption by a square-barrier potential\footnote{Total transmission is used often to describe this phenomena. Since we are interested in exploring stars as well, we prefer to keep the same object-focused nomenclature everywhere. In other words, from the point of view of a {\it potential barrier} there are real frequencies for which a monochromatic wave simply goes through without any reflection. This is the total transmission mode. If, however, we consider the physical system creating the barrier as the entire half plane to the left of a barrier, then the wave was absorbed. We choose here the latter.}, where destructive interference suppresses reflection entirely at specific energies~\cite{Merzbacher:1997}.
A similar stationary-wave approach has been applied to study wave propagation in curved spacetimes, particularly around compact stars and black holes (BHs)~\cite{Futterman:1988ni,Dolan:2008kf,Pijnenburg:2022pug}. Despite their name, the scattering of monochromatic waves off BHs never gives rise to total absorption~\cite{Teukolsky:1974yv,Chandrasekhar:1985kt,Brito:2015oca}\footnote{As we will make clear, BHs do have {\it virtual} absorption (VA) modes. In our terminology, total absorption is reserved for {\it real} frequencies. BHs do have modes for which there is no reflection, but these lie in the complex plane; we call them virtual absorption modes.}. The fundamental reason lies in curved spacetime physics. Although BHs are intrinsically \textit{dissipative} systems (they absorb energy at the horizon and radiate away to infinity), the warping of spacetime and the centrifugal barrier cause a certain amount of reflection near the light ring. 
Furthermore, their dissipative nature means that they cannot support any stationary eigen-states~\cite{Berti:2009kk,Cardoso:2019rvt,Berti:2025hly}. Thus, while BHs do admit total absorption modes, these lie far in the complex plane~\cite{MaassenvandenBrink:2000iwh,Cook:2016ngj}, inaccessible to traditional scattering experiments.
Recently, however, a striking phenomenon has been identified in material science~\cite{Baranov:17,doi:10.1126/sciadv.aaw3255,https://doi.org/10.1002/advs.202301811,2025JAP...137w4701M,2025arXiv250603485F}. By carefully tailoring the initial conditions of an incoming wavepacket (most notably its shape and frequency), no longer monochromatic in the usual sense, it is possible to achieve \textit{coherent perfect absorption}, with the incoming wave fully absorbed by the medium with no reflection or transmission. This effect has been observed in a variety of physical systems -- including optical, acoustic, and mechanical systems --  and shows that reflection and absorption can be completely suppressed even in lossless media, allowing energy to be stored and later released through precise control of the wavepacket's properties.
This raises a natural and intriguing question:~is coherent perfect absorption limited to engineered materials, or is it a more universal phenomenon that can also occur in BH spacetimes? More specifically, is it possible to design wavepackets that are \textit{perfectly absorbed} by a BH, in direct analogy with coherent absorption in materials? 

To answer these questions, this work presents comprehensive time- and frequency-domain studies of scattering in toy-model setups, but also in the background of compact objects, including BH spacetimes. 
For time domain studies, we develop careful strategies to fine tune the initial data for the effect to become apparent in our simulations.
Moreover, exploiting a modern infrastructure for mode analysis in BH physics resulting from conformal geometry and the hyperboloidal framework~\cite{Zenginoglu:2011jz,PanossoMacedo:2023qzp}, we introduce new techniques to calculate total virtual absorption (VA) modes in BH spacetimes (these are traditionally called TTM modes in the literature concerning BH physics; for the reasons explained we will refer to them here as total VA modes). Apart from reproducing the well-known values associated with the algebraically special modes in $4$-dimensional BH spacetimes, we uncover a whole new family of VA modes in higher-dimensional spacetimes. Our findings point to new pathways for energy accumulation in curved geometries, with potential implications for BH formation and instabilities akin to BH bombs~\cite{Press:1972zz,Cardoso:2004nk}.
This paper is organised as follows. We start in Sections~\ref{sec:TA} and~\ref{sec:TVA} by explaining the concept of total absorption and total virtual absorption in several toy models. We then proceed by understanding these phenomena in the context of ultracompact objects (Section~\ref{sec:TVAUC}) and BHs (Section~\ref{sec:TVABH}). We end with discussing the implications of these findings in Section~\ref{sec:discussion}.
\section{Total absorption}\label{sec:TA}
After separation of variables using spin-$s$ harmonics $_{s}Y_{\ell m}$, massless waves around non-spinning stars and BHs are governed by the master equation~\cite{Berti:2009kk},
\begin{equation}
\begin{aligned}
&\frac{\partial^2 \psi}{\partial t^2} - \frac{\partial^2 \psi}{\partial x^2}+ V_s(x) \psi = 0\,, \label{eq:wave_final}
\end{aligned}
\end{equation}
where the effective potential $V_s \to 0$ when $x\to \pm \infty$, depends on the geometry under consideration and on the spin of the massless field, $s=0,1,2$ for scalar, vector and tensor fields respectively. The tortoise coordinate $x$ spans the entire real axis. Note that the BH horizon is at $x=-\infty$.

Our aim is to explore how gravitationally-bound systems respond to external incoming wavepackets of radiation with a non-trivial time-dependence, but it is clear from the above that the determining feature is the structure of the potential $V_s$. Therefore, before dealing with curved backgrounds, it is useful to consider a simpler, well-understood system:~a rectangular barrier potential,
\begin{equation}
V(x) =
\begin{cases}
V_0\,, & \text{for } -L < x < L \\
0\,,   & \text{elsewhere}\,.
\end{cases}
\label{eq:constant_potential}     
\end{equation}

Consider evolving problem~\eqref{eq:wave_final} with initial data (ID) of Gaussian form,
\begin{equation}
\psi(0, x) = e^{-i\Omega_{0} x} \, e^{-\frac{(x - x_{0})^2}{\sigma^2}} \,,\quad
\partial_t \psi = \partial_{x} \psi \,,\label{initial_conditions_I}
\end{equation}
representing a pulse traveling towards the potential barrier with frequency $\Omega_0$. For $\Omega_0 \sigma \lesssim 1$, the ID corresponds to a compact pulse without a well-defined frequency; for $\Omega_0 \sigma \gg 1$, many wavelengths ``fit'' within the width $\sigma$ of the pulse, yielding a nearly monochromatic wave.

\begin{figure}
\includegraphics[width=0.5\textwidth]{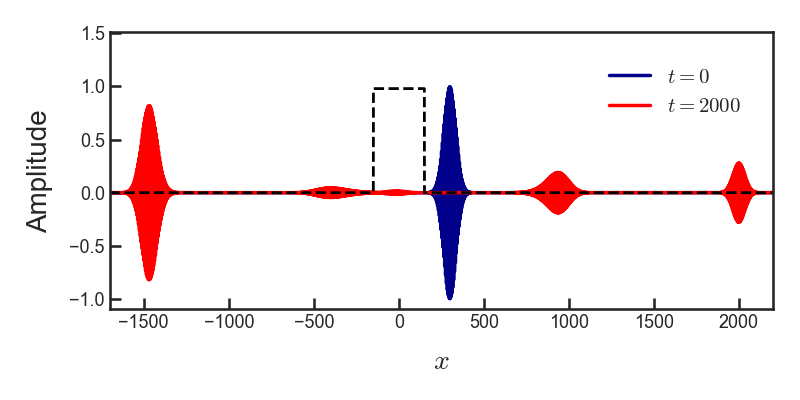}
\caption{Scattering of a Gaussian pulse off a rectangular barrier potential (black dashed lines) with $L=150$ and $V_0=0.980$~\eqref{eq:constant_potential}. The initial pulse (Eq.~\eqref{initial_conditions_I} with $\sigma=50, \Omega_0=1.2$ and $x_0=300$) approaches the barrier from the right. Red lines denote pulses traveling away from the barrier. The scattering of this pulse gives rise to transmitted and reflected components, and ``sluggish returns'' of waves with a small group velocity inside the barrier than eventually leak out.}\label{fig:IDA11_Overplot}
\end{figure}

Figure~\ref{fig:IDA11_Overplot} shows the evolution of Gaussian ID~\eqref{initial_conditions_I} with $\sigma=50,\Omega_0=1.2$ and $x_0=300$ scattering off a barrier with $L=150$ and $V_0=0.98$.
Clearly, upon interacting with either side of the barrier, part of the wave is transmitted and part is reflected. Thus, the initial pulse ends up being partitioned in several copies of itself. The time delay between such ``sluggish returns'' is approximately twice the travel time inside the barrier. With a group velocity $v_{\rm g}=\sqrt{1-V_0/\Omega_0^2}$ and for these parameters, we expect a delay of $\sim 1062$; in good agreement with the observed delay in Fig.~\ref{fig:IDA11_Overplot}. If the ID width increases beyond $2L/v_g$, the returns disappear and the reflected and transmitted components smoothly merge onto a single exponentially decaying signal described by one or more of the modes of the barrier. The requirements for there to be a well defined frequency in the ID is that $\sigma>2\pi/\Omega_0$, but by tuning $\Omega_0$ one can always satisfy the prior constraint. Note that these ``returns'' are ID dependent and not related in any way to characteristic modes of the system.

Although we used a time-domain evolution, the scattering problem is usually approached via monochromatic, constant amplitude waves~\cite{Teukolsky:1974yv,Brito:2015oca,CoG}. By assuming an harmonic dependence for the field, $\psi(t, x) = e^{-i\omega t} R(x)$, one is led to a second-order ordinary differential equation for $R$, which is straightforward to integrate. The regular, ingoing solution to the left $R\sim e^{-i\omega x},\, x\to-\infty$ has the following behavior at large distances,
\begin{equation}
R \to A_{\rm in}e^{-i\omega x}+A_{\rm out}e^{i\omega x}\,.\label{scattering_behavior}
\end{equation}

\begin{figure}
\includegraphics[width=0.5\textwidth]{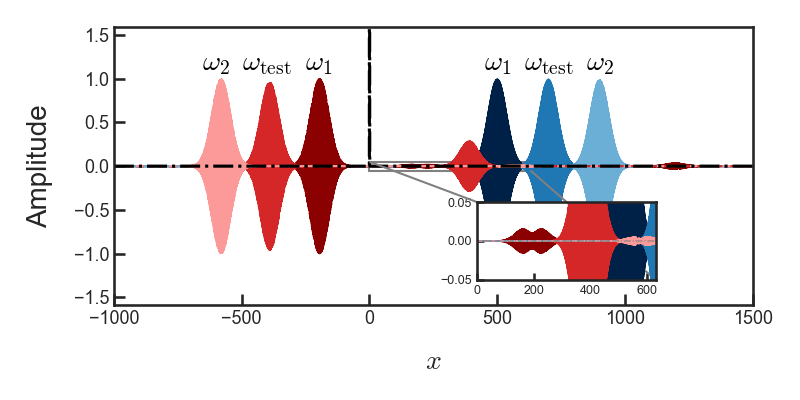}
\caption{Scattering of nearly monochromatic pulses off a rectangular barrier. Incoming Gaussian wave packets have frequency equal to total absorption mode $\Omega_0=\omega_1$ (dark blue, $\omega_1\simeq 5.09$, cf. Eq.~\eqref{eq:omega_levels}), IDA2 (blue, $\Omega_0=\omega_{\rm test}=6$), and $\Omega_0=\omega_2$ (light blue, $\omega_2 \simeq  7.45$). Outgoing waves after barrier-passage are shown in dark red ($t=700$), red ($t=1100$), and light red ($t=1300$), respectively. The inset zooms in on the reflected waves, illustrating that frequencies~\eqref{eq:omega_levels} indeed give rise to {\it total absorption}, while the intermediate value $\omega_{\rm test}$ does not. Moreover, the reflection amplitude decreases as $n$ increases in Eq.~\eqref{eq:omega_levels}.}\label{fig:Total_Absorbtion}
\end{figure}
For the rectangular barrier, it is a trivial matter to solve exactly for the coefficients $A_{\rm in}, A_{\rm out}$. One finds that total absorption -- $A_{\rm out}=0$ in Eq.~\eqref{scattering_behavior} -- occurs at discrete {\it real} frequencies,
\begin{equation}
\omega_n=\sqrt{\frac{n^2\pi^2}{4L^2}+V_0}\,,\qquad n=1,\,2,...\label{eq:omega_levels}
\end{equation}
Note that these are purely {\it real} frequencies, and should thus manifest in scattering of monochromatic pulses. 

We prepared initial data of family~\eqref{initial_conditions_I} with $\sigma=50$, and $\Omega_0=\omega_1$ (with $x_0=500$) and $\Omega_0=\omega_2$ (with $x_0=900$) to understand their scattering properties. We used $L=0.5$ and $V_0=16$. A test run with $\Omega_0=6$, $r_0=700$ allows to understand how special $\omega_n$ are among the parameter space. Our results are shown in Fig.~\ref{fig:Total_Absorbtion}. As predicted, $\omega_1$ and $\omega_2$ exhibit nearly total absorption, while the intermediate $\omega_{\rm test}$ does not. The inset highlights the reflected components, whose amplitude decreases with increasing $n$ in Eq.~\eqref{eq:omega_levels}. It's telling that the total absorption frequencies satisfy $\omega_n^2>V_0$, and are real.
\section{Total virtual absorption by a barrier potential}\label{sec:TVA}
%
\begin{figure}[ht!]
\includegraphics[width=0.5\textwidth]{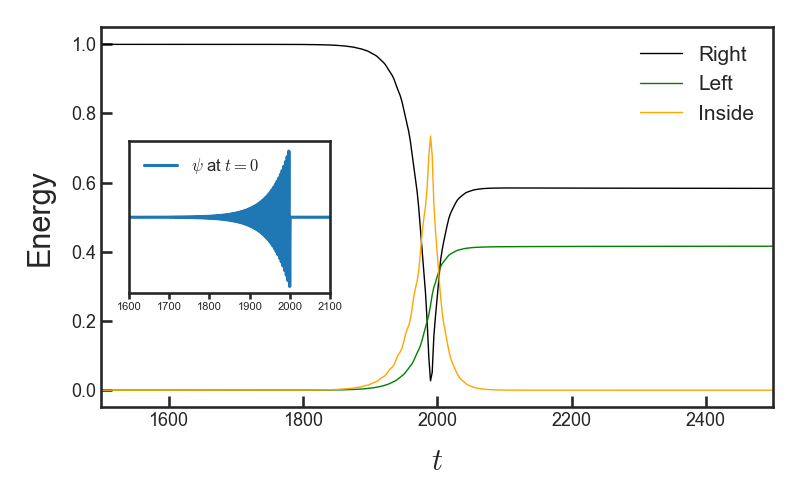}
\caption{Energy content of a region where a double-barrier is placed. Initially a virtual absorption mode \eqref{VA_double}--\eqref{initial_conditions_II} is hitting the barrier from the right. For as long as this mode is exciting the barrier, it is absorbed and stored by the system. Thus the total energy to the right of the barrier (black line) decreases, while energy content within (yellow) the barrier increases. The wave packet is transmitted to the left of the barrier (green). When the excitation stops, virtual absorption stops and the signal is sent out of the barrier. Inset shows shape of the initial data.}\label{fig:Virtual_Transmission_Double_Barrier}
\end{figure}
There are however, circumstances where total absorption modes lie in the complex plane. We call these {\it virtual absorption modes}.
Take for example two rectangular barriers, of heights $V_0=12$ between $x=[8.5,9.1]$ and $V_1=9$ between $[10.5,11]$. We find a low-energy mode,
\be
\omega_{\rm VA}=1.513+0.021i\,.\label{VA_double}
\ee

We now excite the system with such a mode. We do this via the following family of initial data, consisting of an exponentially growing sinusoidal wave truncated by a Gaussian at a radius $x_0$ ($\eta\equiv x-x_0$),
\begin{equation}
\psi(0,x)=\left(e^{\Omega_{\rm I}\eta}\Theta(-\eta)+e^{-(\eta/\sigma)^2}\Theta(\eta)\right)\,\cos{[\Omega_{\rm R}\,\eta]}\,.
\label{initial_conditions_II}
\end{equation}
with $\omega_{\rm VA}=\Omega_{\rm R}+i\Omega_{\rm I}$, $x_0=2000$ the truncation point and $\sigma=\sqrt{10}$ the width of the sinusoid.

We compute the energy density
\begin{equation}
\varepsilon(t,x)=\frac{1}{2}\left| \partial_t\psi \right|^2 + \left| \partial_{x}\psi \right|^2 + V\left| \psi \right|^2\,,
\end{equation}
and integrate over $x=[a,b]$ at time=$t$,
\begin{equation}
E_{ab}(t)=\int_{a}^{b}\varepsilon(t,x)\,\mathrm{d}x\,.
\end{equation}
One can then calculate the energy of the ID, and of the transmitted and reflected wave. Our results are summarized in Fig.~\ref{fig:Virtual_Transmission_Double_Barrier}, one of the key messages of this work. We calculated the total energy to right of the barriers ($x>11$) to the left ($x<8.5$) and within. As long as the initial wavepacket is hitting the barrier, it absorbs the incoming energy, filling up the barrier and transmitting to the left until the excitation stops. Then, waves tunnel out of the barrier to the left and to the right of the barrier. It can be seen that the energy of the right region becomes essentially zero at a time close to $t=2000$, corresponding to the end of the exponential sinusoidal front of the wave packet. Then, there is a sharp increase in the energy within the right region, which is highly correlated with the discontinuous nature of the double potential barrier. This is the 'virtual' aspect of the absorption. If one excites the cavity for an infinite amount of time, there would be no reflection. Virtual absorption here will stand for a process whereby a system absorbs radiation while being illuminated, but eventually releases it because it is unable to store radiation permanently. These systems have total absorption modes in the complex plane. This result can also be understood in the frequency domain. Right-moving VA modes (or, as we stated, also called TTM$_R$) are solutions with only ingoing waves at the right of the potential, and only outgoing waves to the left. In the frequency domain, these are infinitely long in space and stationary. Thus, the time domain scattering of real, total absorption modes  (Fig.~\ref{fig:Total_Absorbtion}) results in $100\%$ transmission to the left of the potential; similarly, the time domain scattering of initial data corresponding to a complex VA mode, would yield perfect absorption. Thus, if the time domain initial data constructed in Fig. \ref{fig:Virtual_Transmission_Double_Barrier} were infinitely long in space, it would correspond to the frequency domain solution and there would not be any reflection to the right of the potential. However, as this excitation stops due to the finite length of the wavepacket, these scattering process shows a non negligible amount of reflection after the excitation stops. So, we conclude that this is the virtual aspect of these complex VAs and virtual absorption occurs until the excitation stops.

Exciting the system with a frequency slightly (10\% or more) off \eqref{VA_double} results in much smaller values of trapped energy. Likewise, one can excite the barrier with a packet with $\Omega_{\rm I}$ artificially set to zero. We observe a similar result: the imaginary component keeps the excitation and it hinders any reflection until the signal is cutoff. Indeed, exciting the system with a monochromatic wave ($\Omega_{\rm I}=0$) we see immediate reflection as soon as the incident wave arrives. The wave with VA frequency gives reflection flux only after the excitation ceases.

We tested a number of other systems, with the same outcome: the presence of a real total absorption mode results in actual absorption and therefore transmission through the barrier, even if the mode is low energy. Complex VA modes translate to virtual absorption, whereby the system stores the energy for as long as the excitation is active and then releases it. This applies, in particular, to a purely imaginary VA mode such as so called algebraically special modes. However, these modes have $0$ quality factor (i.e $\Re(\omega)/\Im(\omega)$) which makes time domain simulations hard to construct. In absence of a real part for the characteristic frequency, their propagation properties are also less clear. These modes deserve to be investigated in detail, but they are not within the scope of this work.

The phenomenon is general: resonances in the complex plane can be assessed with carefully crafted ID. To emulate this phenomena in the case of a single barrier potential, we need to modify slightly the experiment, and also 
stimulate a single barrier from {\it both} directions in a finely tuned way. This strategy results in adapting the original boundary conditions to impose 
only ingoing waves from both sides of the barrier. The resulting modes, thus, correspond to the complex conjugate of the characteristic QNMs of the system.

\begin{equation}
2i\omega_{\rm VA}\sqrt{\omega_{\rm VA}^2-V_0}=-V_0\sin\left(2\sqrt{\omega_{\rm VA}^2-V_0}L\right)\,.
\label{eq:QNM_VA}
\end{equation}
There are an infinity of solutions to this equation for any fixed $L,\,V_0$. For $L=1$, $V_0=16$ for example, we find one solution at $\omega_{\rm VA}=4.25316 + 0.12731i$.

\begin{figure}[ht!]
\includegraphics[width=0.5\textwidth]{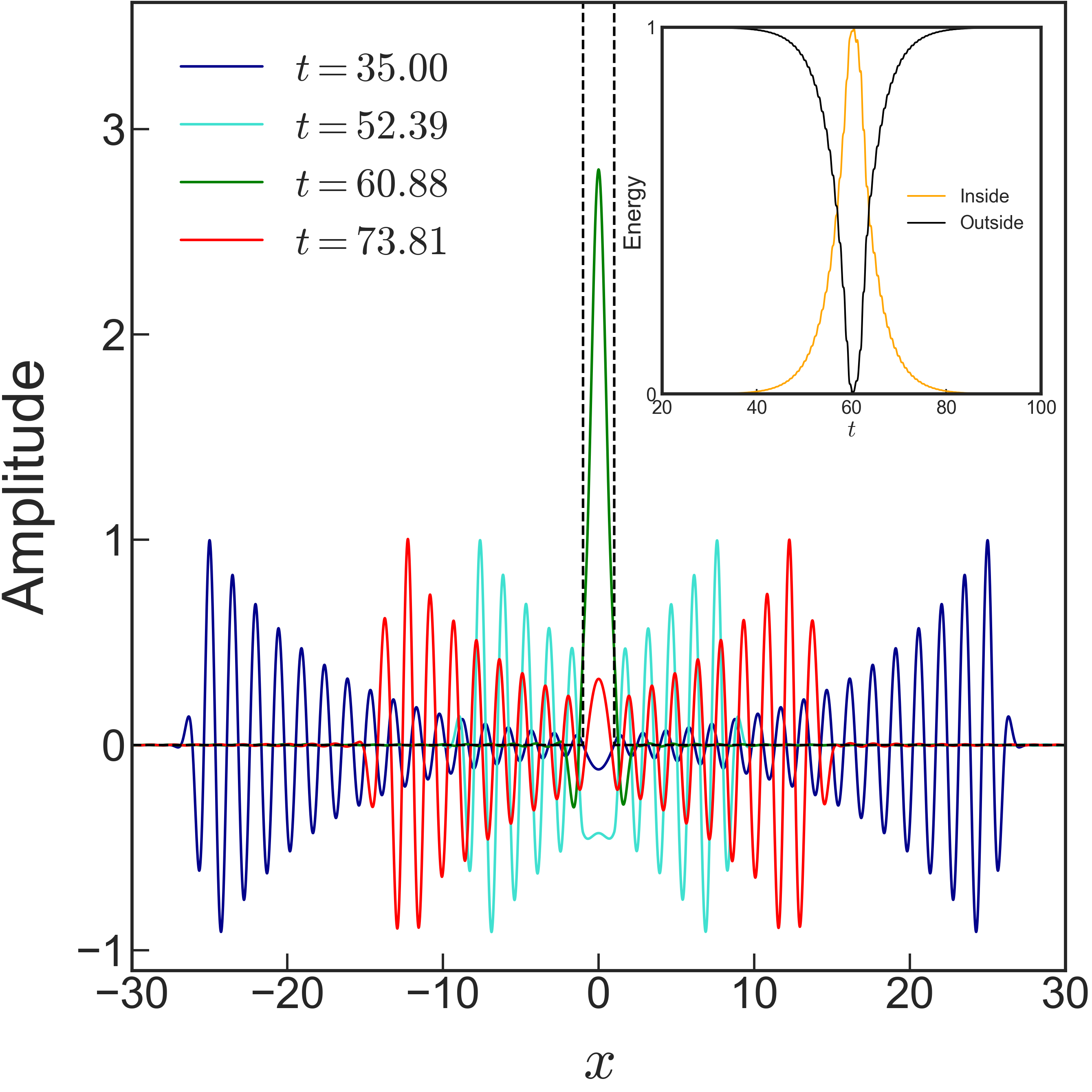}
\caption{Scattering off a rectangular barrier, leading to perfect two-sided virtual absorption. An incoming wave ($t=35$, dark blue) moves towards the barrier on both sides. It travels at the speed of light and at $t= 52.39$ (light blue) nearly half of it penetrated the barrier where it got stored. At $t= 60.88$ (green) all the incoming packet is stored within the barrier, which is now ready to release the energy content. At $t=73.81$ (red) most of the energy is moving away from the barrier on both sides. 
The inset shows the energy stored inside (orange) and outside (black) the rectangular barrier. At $t\approx 61$ all the energy is contained within the barrier (showing unitary storage). Afterwards, the stored energy decays exponentially, and is well described by $Ae^{-\alpha t}$ with $\alpha=0.25902$ twice the characteristic frequency $\omega_{\rm I}$.
}\label{fig:Virtual_Absorbtion_Over_Plot}
\end{figure}
We excite the system with such a mode, which in essence amounts to following back in time the relaxation of an excited barrier. We send a wave from the right of the potential with $x_0=60,\, \sigma=1$ for the ID~\eqref{initial_conditions_I} {\it and} its mirror reflection with respect to the $x=0$ vertical line, both going towards the potential barrier. Our results are summarized in Fig.~\ref{fig:Virtual_Absorbtion_Over_Plot}. The two incident waves meet at the barrier and are stored {\it with nearly unit efficiency} while the barrier is being ``bombarded''.
We have created a perfect absorber. At $t \approx 61$, all the energy is stored inside the barrier (see inset), marking perfect storage. Afterwards, the stored energy leaks out exponentially, via ringdown; indeed, we find that the flux decays exponentially afterwards, well fit by $\sim e^{-0.259 t}$, as one would expect (i.e., twice the imaginary part of the QNM~\eqref{eq:QNM_VA}).

The results summarized in Figs.~\ref{fig:Virtual_Transmission_Double_Barrier} and~\ref{fig:Virtual_Absorbtion_Over_Plot} allow one conclude the following: a two-sided excitation of any potential by the complex conjugate of its QNMs and one sided excitation by the complex VAs result in \textit{Virtual Absorption}. Thus, scattering of QNMs and TTMs are two different realizations of one physical phenomenon, virtual absorption. It follos also that VA scattering occurs for any potential that admits QNMs and/or TTMs.

\section{Total Virtual absorption of ultracompact objects}\label{sec:TVAUC}
The above establishes that the complex total absorption modes of any system can be accessed with time-dependent signals, modulated to match the time dependence of the relevant modes. We now calculate how this takes place in compact stars. For simplicity here (because we are eventually interested in approaching the BH limit), we take an exterior Schwarzschild geometry,
\begin{equation}
ds^2=-f\mathrm{d}t^2+\frac{\mathrm{d}r^2}{f}+r^2\mathrm{d}\Omega^2\,,
\label{eq:metric}
\end{equation}
with $f=1-2M/r$, and the tortoise coordinate defined by $\mathrm{d}x/\mathrm{d}r=1/f$. The interior will not be very relevant to us, and we replace it with Dirichlet boundary conditions at $r_0=2M(1+\epsilon)$~\cite{Cardoso:2016rao,Cardoso:2016oxy,Cardoso:2017cqb,Cardoso:2019rvt,Boyanov:2022ark}. The QNMs of this spacetime and setup are well studied~\cite{Cardoso:2016rao,Cardoso:2016oxy,Cardoso:2017cqb,Cardoso:2019rvt,Maggio:2018ivz,Boyanov:2022ark}. The effective potential for massless waves with angular momentum $\ell$ is 
\be
V_s=\left(1-\frac{2M}{r}\right)\left(\frac{\ell(\ell+1)}{r^2}+(1-s^2)\frac{2M}{r^3}\right)\,,  
\ee
Given the boundary conditions imposed at the surface of these objects, it is easy to show that the total absorption or VA modes are complex conjugates of QNMs\footnote{QNMs obey outgoing Sommerfeld condition at large spatial distances. For BHs, QNMs correspond to waves ingoing towards the horizon. By contrast, VA modes correspond to ingoing solutions at large distances. For BHs they correspond to ingoing waves at the horizon.}. Using a direct integration approach, we find different families of them. For example,
\begin{equation}
M\omega_{\rm VA}=0.5122+i0.01643 \,,\qquad \epsilon= 10^{-3}\,,\\
\label{Eq:Omega_for_star}
\end{equation}
for a scalar field and $\ell=2$. This is the virtual absorption result for frequency given by Eq.~\eqref{Eq:Omega_for_star} which creates a Dirichlet boundary at $x_{\rm b}\approx -11.8$ in tortoise coordinates.
\begin{figure}[ht!]
\includegraphics[width=0.5\textwidth]{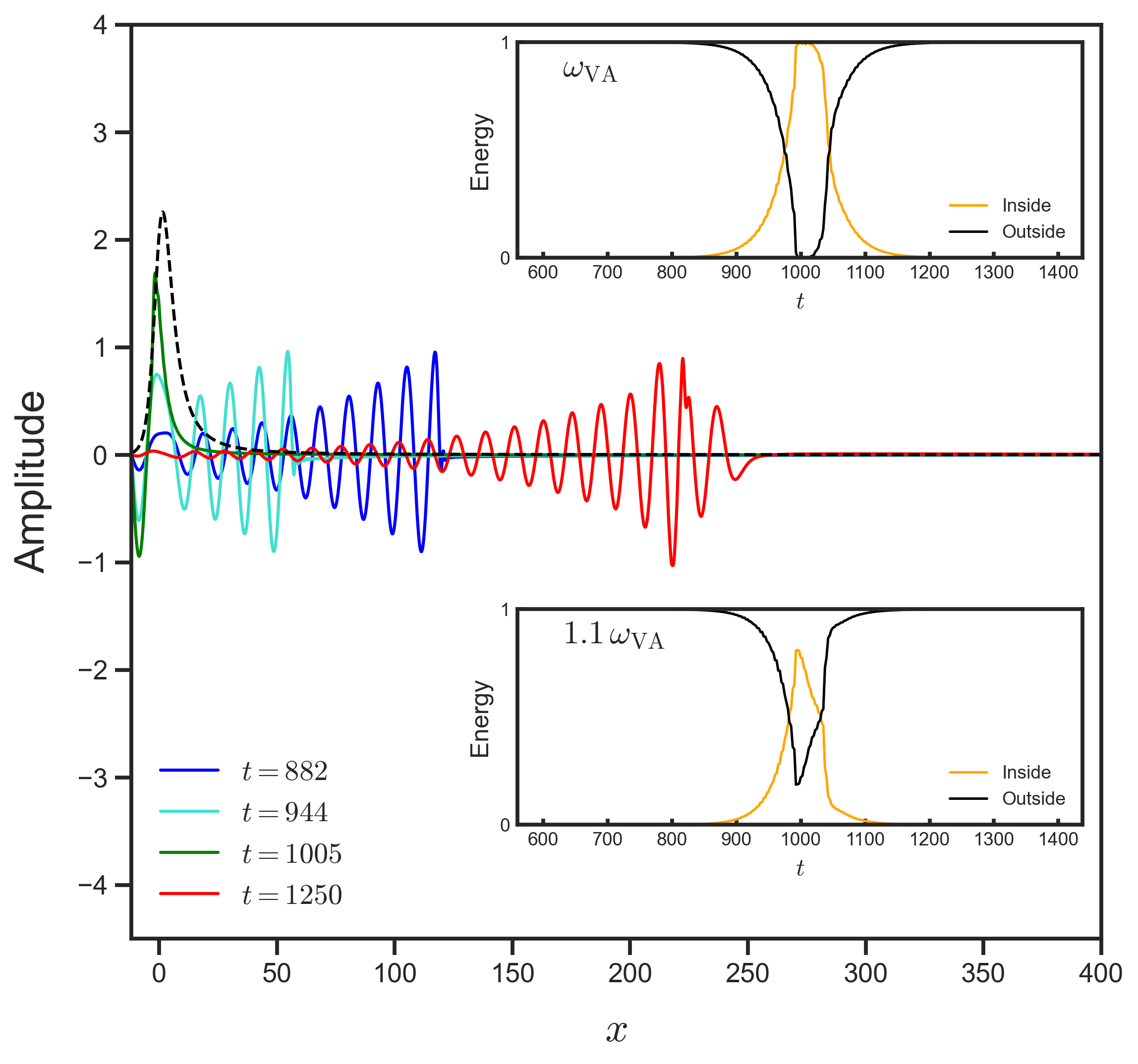}
\caption{Scattering of VA waves off a ultracompact object ($\epsilon=10^{-3}$), leading to perfect virtual absorption. An incoming wave at $t=882$ (blue) moves towards the potential on right side of it. It travels at the speed of light and at $t= 944$ (light blue) nearly half of it penetrated the potential where it got stored. At $t=1005$ (green) all the incoming packet is stored within the star (more precisely, within the potential barrier), which is now ready to release the energy content. At $t=1250$ (red) most of the energy is now out moving away from the barrier on the right side. 
The inset shows the energy stored inside (yellow) and outside (black) the potential. At $t\sim 1000$ all the energy is contained within the cavity (showing unitary storage). We define the inside region as $-11.8\le x\le10.0$ (smoothly varying potentials do not have well-defined boundaries, the inside is loosely defined based on our simulations. Afterwards, the stored energy decays exponentially, and is well described by the ringdown modes of the object. The inset below shows that VA is not an accident: once we change the frequency of the incoming wave by only 10\%, total absorption is no longer observed.
}\label{fig:Virtual_Absorbtion_Stars}
\end{figure}

We then illuminate the object ($\epsilon=10^{-3}$) with the VA mode above. Our results are summarized in Fig.~\ref{fig:Virtual_Absorbtion_Stars}, and follow closely the behavior of the toy model. The star displays virtual absorption for as long as a VA mode impinges. The radiation gets enclosed in a region of a spatial extent $\sim 3M$ dictated by the effective potential barrier. Total absorption is seen in the inset (top right) of Fig.~\ref{fig:Virtual_Absorbtion_Stars}. To show that we are indeed catering to a resonance in the complex plane, the inset below shows the amount of absorption once we slightly change the parameters of the initial data $(\Omega_{\rm R}, \Omega_{\rm I}) \to 1.1 (\Omega_{\rm R}, \Omega_{\rm I})$. Imperfect absorption is now visible. We find that total absorption indeed requires a fine tuning of ID with the VA resonances of the object.

One might expect higher frequencies to result in more transmission; our results show that this is not true close to $\omega \sim \omega_{\rm VA}$. This means that this specific frequency shows a discrete behaviour in the spectrum, evidence of VA. We also tested changing initial data as $(\Omega_{\rm R}, \Omega_{\rm I}) \to 0.9 (\Omega_{\rm R}, \Omega_{\rm I})$. This change also results in imperfect absorption. In this scenario, the effective width of the incident wave is approximately $250$ whereas the width of the inside region is $21.8$. Thus, the effect we report can not be caused by any artifact based on relative width sizes. Also one can realize that the peak of the energy trapped inside the potential stays essentially zero for some time; this duration is determined by the scale of the potential, and is also apparent in Fig.~\ref{fig:Virtual_Absorbtion_Over_Plot} which has a smaller inside region (because of the nature of the potential). VA is associated with energy accumulation and lack of reflection during the interaction, thus the duration of the inside energy peak is a result of the shape of the potential.

When the cavity is made larger, i.e., when $\epsilon$ decreases, the width of initial data required to see perfect absorption {\it increases}:~the incoming pulse must interact with the boundary to be ``aware'' of the necessary boundary conditions. In fact, in the limit that $\epsilon \to 0$, which corresponds to the BH limit, total absorption is impossible, as it requires an infinite train of radiation.\footnote{For ultracompact objects, VA modes are obtained by solving the time-independent wave equation with a Dirichlet boundary condition. The reflection produced by this boundary is essential for the mode to exhibit VA, and in the BH limit an infinite train of radiation is required for the wave to reach the boundary. This is not the case for BH VA modes, where there is no reflective boundary and the event horizon imposes a purely ingoing condition.}
\section{Total virtual absorption of black holes}\label{sec:TVABH}

In this section we demonstrate VA in BH geometries, starting with two-sided VA and then demonstrating one-sided VA. The basic strategy follows that of the toy models discussed earlier:~for two-sided VA, we construct time-domain wave packets from frequency-domain solutions (in practice, complex conjugates of QNMs) and study their scattering off the effective BH potential. When the incoming packet is tuned to an appropriate complex frequency, the field energy becomes temporarily stored in the potential region and is subsequently re-emitted once the excitation ends. For one-sided VA we will first find VA modes of higher dimensional BHs (reported here for the first time) and then ``illuminate'' BHs with the corresponding initial data.

\subsection{Two-sided VA: Schwarzschild and Bardeen cases}
%
\begin{figure}[ht!]
\includegraphics[width=0.5\textwidth]{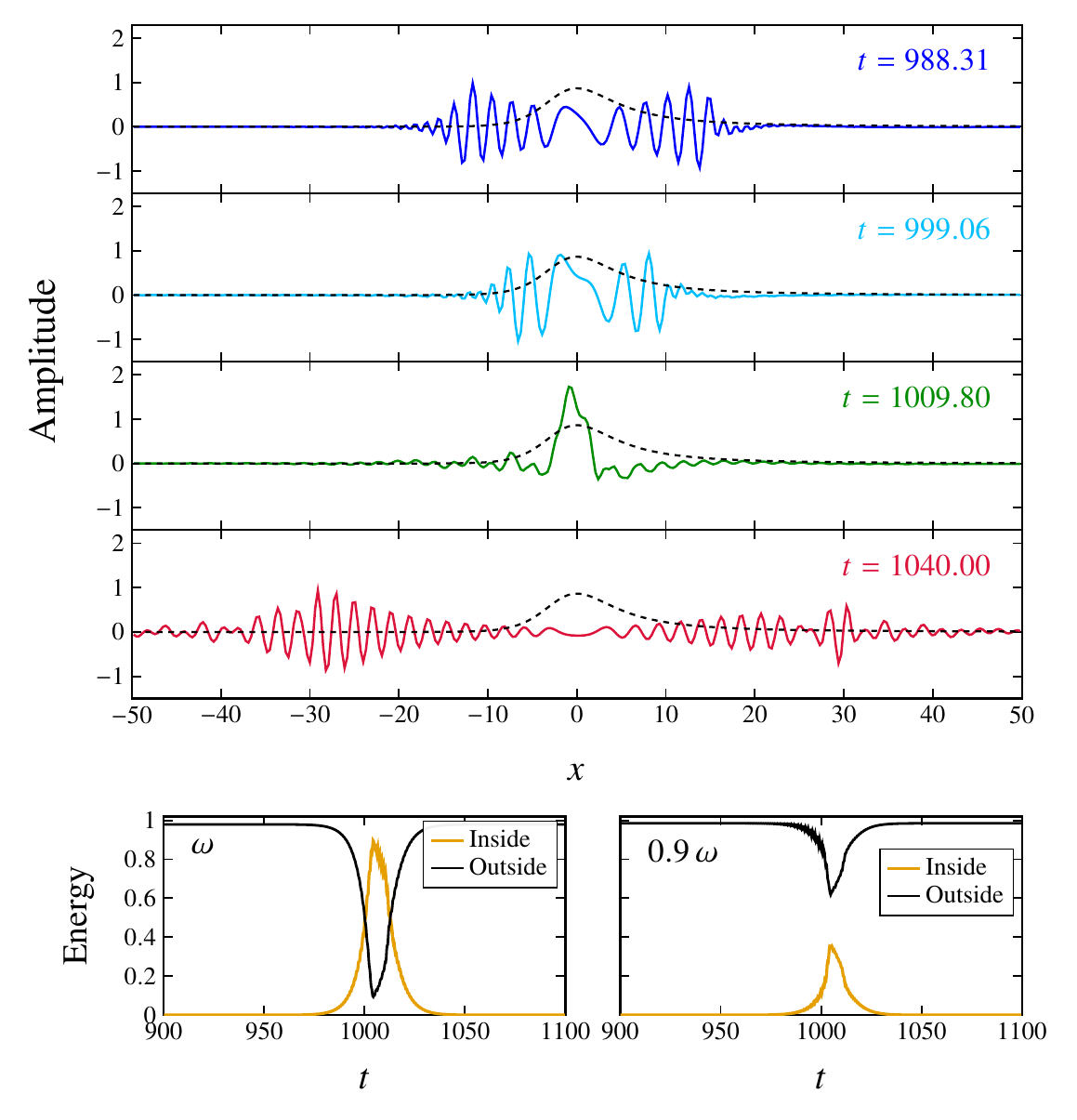}
\caption{Two-sided VA scattering in Schwarzschild geometry. The initial VAM has $\omega_{\rm VA}M=2.983+0.0962i$ and is incident from the right; its mirror image is incident from the left. We define \textbf{Outside: $x<-2$ and $x>3$} and \textbf{Inside: $-2<x<3$}. \textbf{Top:} Curves at $t=989$ (blue) and $t=995$ (light blue) show incoming waves; $t=1005$ (green) shows the moment of VA; $t=1040$ (red) shows the outgoing wave after the excitation ends. \textbf{Bottom:} energy versus time for $\omega_{\rm VA}$ (left) and $0.9\omega_{\rm VA}$ (right). The black curve is the outside energy and the yellow curve the inside energy. While the mode excites the potential the outside energy drops and the inside energy increases; when the excitation ends the stored energy is re-emitted.}
\label{fig:Virtual_Absorption_Schwarzschild}
\end{figure}
We first construct two-sided VA modes using complex conjugates of well-known QNMs. For a Schwarzschild geometry, we use the $\ell=15$, $n=0$ mode of a scalar field, with frequency $\omega M = 2.983 - 0.0962\,i$ \cite{CoG,Berti:2025hly,Pan:2006ti} (the exact choice of mode is arbitrary, but we do need a high quality factor). The initial data is given by Eq.~\eqref{initial_conditions_II} with $\sigma=1$. One wave packet is launched from the right ($x_0=+1000$), while the second packet is its mirror image with respect to $x=0$ (at $x_0=-1000$), so that both packets carry the complex-conjugate frequency. The resulting scattering is shown in Fig.~\ref{fig:Virtual_Absorption_Schwarzschild}. Here, the effective width of the wave packet is approximately $60$, whereas the width of the inside region is $\sim 5$. Thus, the wave extent is larger than the scattering region.
\begin{figure}[ht!]
\includegraphics[width=0.5\textwidth]{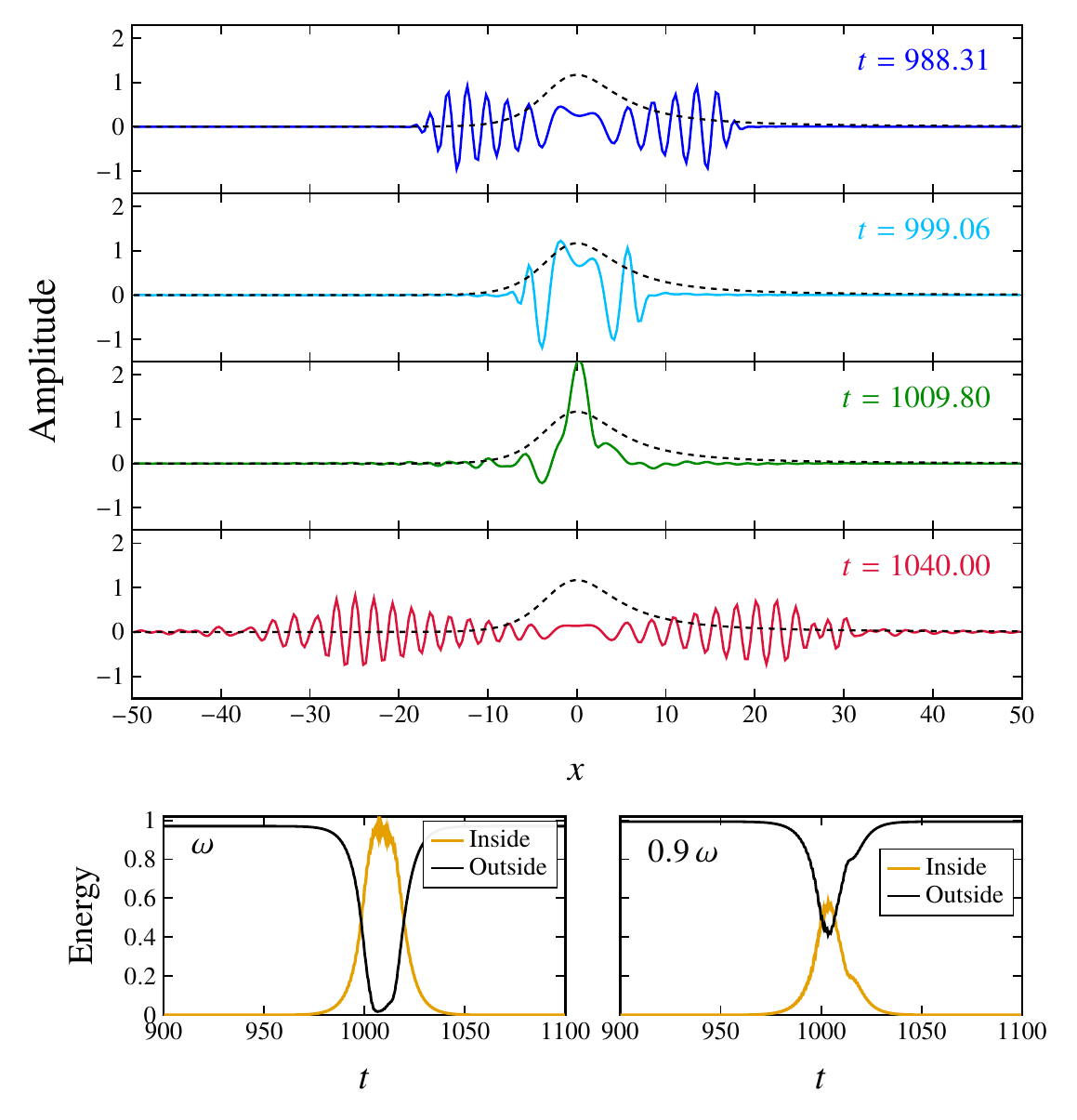 }
\caption{Two-sided VA scattering in a Bardeen geometry. The geometry has a VA mode at $\omega_{\rm VA}M=2.924 + 0.09252i$, which we use to produce scattering ID, incident from the right; its mirror image is incident from the left. We define \textbf{Outside: $x<-3.9$ and $x>3.3$} and \textbf{Inside: $-3.9<x<3.3$}. \textbf{Top:} Curves at $t\approx988$ (blue) and $t\approx999$ (light blue) show incoming waves; $t\approx1010$ (green) shows the moment of VA; $t=1040$ (red) shows the outgoing wave after the excitation ends.\textbf{Bottom:} energy versus time for $\omega_{\rm VA}$ (left) and $0.9\omega_{\rm VA}$ (right). The black curve is the outside energy and the yellow curve the inside energy. While the mode excites the potential the outside energy drops and the inside energy increases; when the excitation ends the stored energy is re-emitted.}\label{fig:Virtual_Absorption_Bardeen}
\end{figure}
A similar construction applies to the regular Bardeen BH. Using the metric function
$f(r)=1-2Mr^2/(r^2+q^2)^{3/2}$ with $M=1$ and $q=0.5$, and choosing the $\ell=14$, $n=0$ QNM with
$\omega M=2.92388 -0.0925168i$ \cite{Fernando:2012yw}, we again construct mirrored incoming packets with $\sigma=1$. The resulting scattering, shown in Fig.~\ref{fig:Virtual_Absorption_Bardeen}, exhibits the same VA phenomenology as in the Schwarzschild case. ft{Again, the effective width of the wave packet is approximately $60$ is larger than the width $\sim7.2$ of the inside region.}

Figures~\ref{fig:Virtual_Absorption_Schwarzschild} and \ref{fig:Virtual_Absorption_Bardeen} show that two-sided VA modes in BH potentials reproduce the same qualitative behavior observed in the potential barrier toy model whose result is given in Fig~\ref{fig:Virtual_Absorbtion_Over_Plot}. While the mode excites the potential, energy accumulates in the ``inside region'' and the outside energy decreases; once the excitation stops, the stored energy is released. The shape of the focused wave differs from the toy-model case because of the greater asymmetry of BH potentials.

\begin{figure}[ht!]
\includegraphics[width=0.5\textwidth]{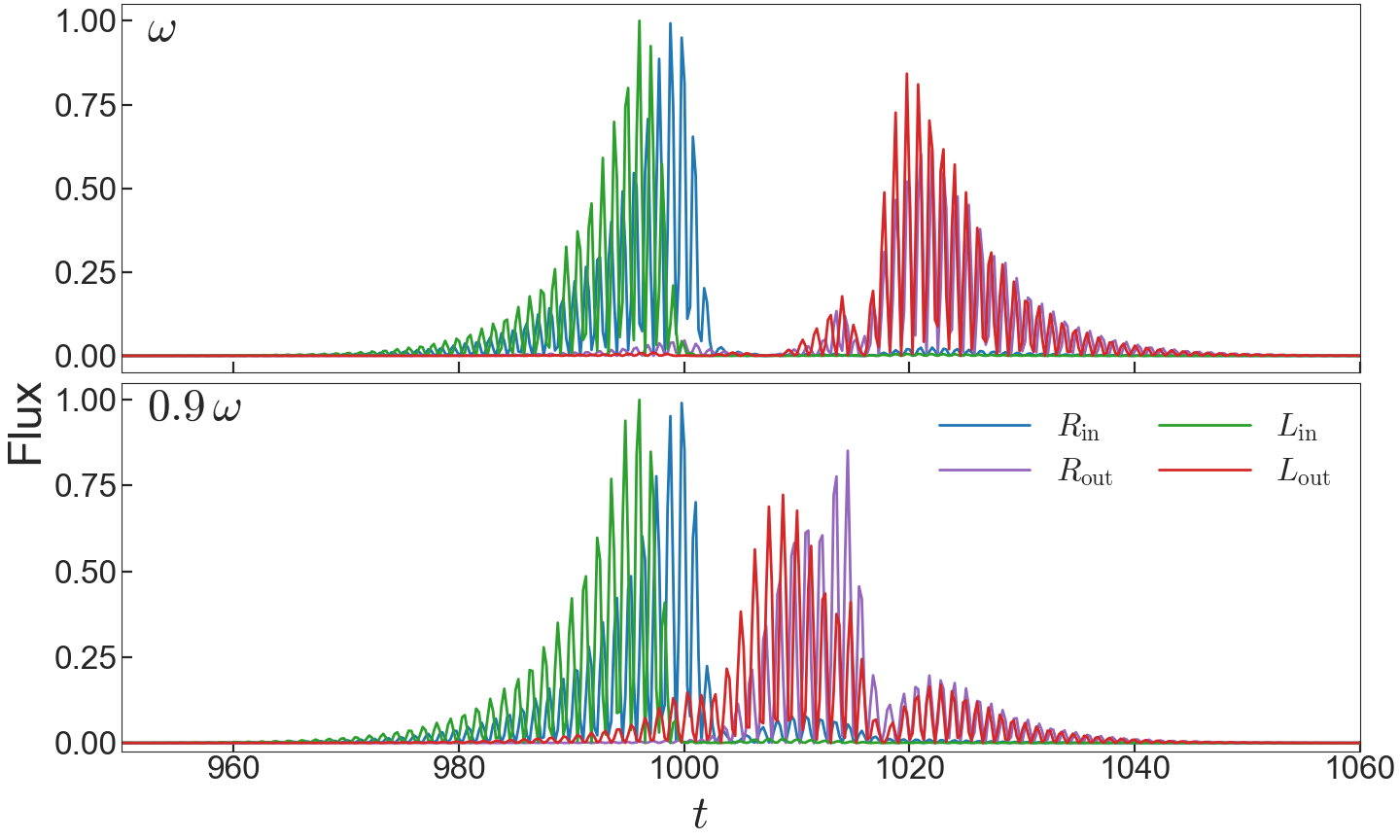}
\caption{Instantaneous flux for the two-sided VA scattering on a Bardeen BH, corresponding to Fig.~\ref{fig:Virtual_Absorption_Bardeen}. Top and bottom panels show fluxes corresponding to two-sided scattering of pulses with frequency $\omega_{\rm VA}$ and $0.9\omega_{\rm VA}$, respectively. The incident wave from the right of the potential and the reflected wave towards the right of the potential are represented by $R_{\rm in}$ (blue) and $R_{\rm out}$ (purple), respectively. Same applies to the left region.
}\label{fig:Virtual_Absorption_Bardeen_Flux}
\end{figure}

It is also possible to see the delay in the reflected waves by looking at the instantaneous flux plots, shown in Fig.~\ref{fig:Virtual_Absorption_Bardeen_Flux}. Here the field is extracted at $x=-5.9$ (left) and at $x=5.3$ (right).
Here we see that the QNMs that we utilized are absorbed for a larger amount of time compared to $0.9\omega_{\rm VA}$. Thus it becomes obvious that using symmetric incoming waves from two sides is a good enough choice despite the asymmetric form of the black hole potential. The only thing that can be seen is a small and earlier reflection for the $\omega_{\rm VA}$ scattering which is caused by dispersion and the asymmetric shape of the black hole potential. However, it can be clearly seen that most portion of the reflected waves are delayed and there is no significant reflection during the accumulation of right and left incident waves into the inside region. Moreover, as the effective width of the incident waves are approximately $10$ times the width of the inside regions, it is unlikely that asymmetric shape of the potential have a huge effect on the VA phenomenon. Also this result shows the equivalence of the vanishing outside energy and delay in the instantaneous flux of the reflected waves.

In a scattering experiment the physical scenario is to have only right incident waves, as a two sided example would require a left incident wave. However this problem can be overcome by defining an observer at the left of the black hole potential and emitting a left incident wave towards the potential.
\subsection{One-sided VA: high-dimensional Tangherlini example}
Explicit studies of VA modes in BH spacetimes are scarce~\cite{MaassenvandenBrink:2000iwh,Cook:2016ngj,Cook:2022kbb,Grozdanov:2023txs,Grozdanov:2025ner}. Motivated by this, we carried out an independent search for VA modes in arbitrary $d$-dimensional BH spacetimes~\cite{Tangherlini:1963bw}. Our analysis is based on a well-established perturbation formalism~\cite{Kodama:2003jz,Kodama:2003kk,Cardoso:2003vt,Matyjasek:2021xfg} and a conformal framework~\cite{Zenginoglu:2011jz,PanossoMacedo:2023qzp}, via a spectral method based on Chebyshev collocation point~\cite{Jaramillo:2020tuu}. We complement this with a continued fraction implementation~\cite{Cardoso:2003vt,Matyjasek:2021xfg} and an analytic large-$d$ approximation~\cite{Emparan:2014cia} (see Appendix~\ref{app:larged} for more details). For the typical resolutions used, all methods agree to about ten decimal places. This search reveals a nontrivial and structured family of complex solutions corresponding to VA modes, showing that BHs are also prone to VA. The qualitative features closely mirror those discussed earlier for simpler systems.

\begin{figure}[ht!]
\includegraphics[width=0.5\textwidth]{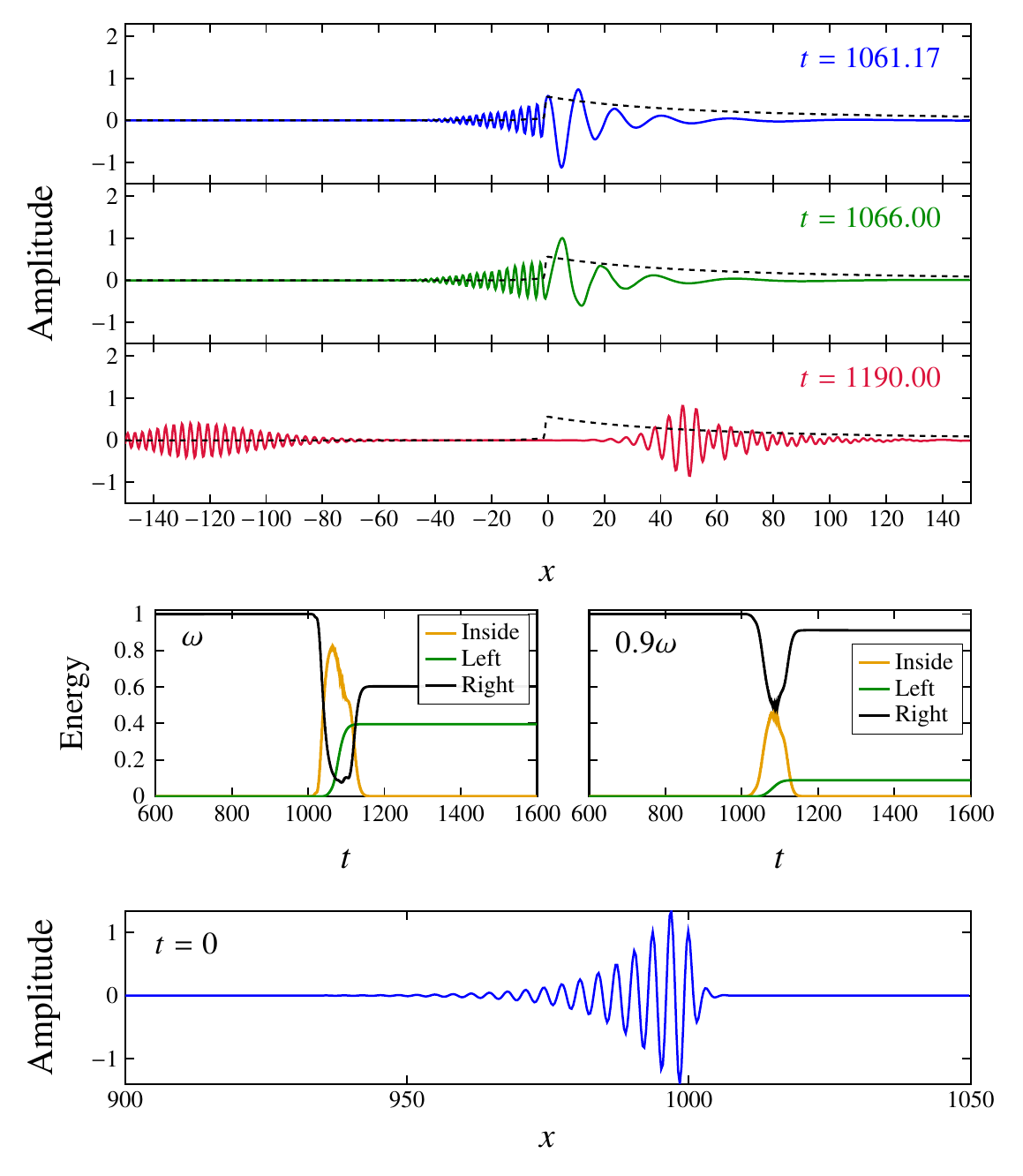}
\caption{Scattering of a scalar VA mode off a $d=400$ Tangherlini BH. \textbf{Top:} The curve at $t=1061$ (light blue) shows the incoming wave, $t=1066$ (green) corresponds to the moment of VA, and $t=1190$ (red) shows the wave after scattering. \textbf{Middle:} The initial wave packet has a small tail.\textbf{Bottom:} energy evolutions for $\omega_{\rm VA}$ (left) and $0.9\omega_{\rm VA}$ (right): black corresponds to the right region, orange to the inside region, and green to the left region. While the mode excites the potential, energy accumulates inside and decreases on the right; when the excitation stops, the stored energy is released and a partial transmission to the left is observed.}
\label{fig:Virtual_Transmission_Tangherlini}
\end{figure}

To demonstrate a clear one-sided VA example in the time domain, we focus on a high-dimensional Tangherlini BH. Figure~\ref{fig:Virtual_Transmission_Tangherlini} shows the scattering of a scalar VA mode off a $d=400$ Tangherlini BH. The VA mode has frequency $\omega_{\rm VA} r_h = 195.1821 + 10.6846\,i$ and we choose $r_h = 100$ to control the exponential growth of the wave during evolution. The left, right, and inside regions are defined according to the shape of the effective potential: \textbf{Left: $x<-14.1$}, \textbf{Right: $x>12.1$}, and \textbf{Inside: $-14.1<x<12.1$}. The left boundary is chosen such that the potential is negligible there. The right boundary can be chosen in different ways; we choose $x=12.1$, as our simulation indicates that most of the pulse is focused in this range. Other choices would result in similar conclusions.

In previous examples, the ratio $\Re(\omega)/\Im(\omega)$ was of order $30$--$70$, which is important for constructing wave packets whose tails carry relatively little energy. As this ratio decreases, suppressing noise from the Gaussian truncation becomes increasingly difficult. This motivates our choice of the $d=400$ VA mode, which has a sufficiently large ratio to allow a clean separation between the exponential front and the tail. The large value of $\Re(\omega_{\rm VA})$ also requires a large horizon radius $r_h$ to avoid numerical blow-up, hence our choice of $r_h=100$.

To further suppress tail contamination, we construct the initial wave packet using a smoother cutoff instead of a sharp step function. Specifically, we replace the $\Theta$ function in Eq.~\eqref{initial_conditions_II} with
\begin{equation}
\psi(0,x)=\left(e^{\Omega_{\rm I}\eta}\left[1-S(\eta;w)\right]+e^{-(\eta/\sigma)^2}S(\eta;w)\right)\cos(\Omega_{\rm R}\eta)\,,
\label{new_initial_condition}
\end{equation}
where $S(\eta;w)=\tfrac{1}{2}\left[1+\tanh(\eta/w)\right]$. To further reduce the tail energy, we superpose this with
\begin{equation}
\psi(0,x)=\left(e^{\Omega_{\rm I}\eta}\left[1-S(\eta;w)\right]-e^{-(\eta/\sigma)^2}S(\eta;w)\right)\cos(\Omega_{\rm R}\eta)\,.
\label{new_initial_condition_2}
\end{equation}
Using $\sigma=\sqrt{1/1000}$ and $w=2$ yields an initial packet whose exponential front carries about $93.5\%$ of the total energy, while the tail carries only $6.5\%$ (see inset in Fig.~\ref{fig:Virtual_Transmission_Tangherlini}).

The scattering shown in Fig.~\ref{fig:Virtual_Transmission_Tangherlini} differs from the two-sided VA observed in Figs.~\ref{fig:Virtual_Absorption_Schwarzschild} and \ref{fig:Virtual_Absorption_Bardeen}. In the two-sided cases, the VA modes correspond to zero-reflection and zero-transmission solutions of the frequency-domain problem, resulting in energy being localized only inside the potential. In contrast, the $d=400$ Tangherlini VA mode is a zero-reflection but nonzero-transmission solution, leading to both a focused wave inside the potential and a transmitted signal to the left. This behaviour is directly analogous to the one-sided VA observed for the double-barrier system in Fig.~\ref{fig:Virtual_Transmission_Double_Barrier}. The focused wave spreads over a wider region due to strong dispersion and the asymmetry of the BH potential, but the essential VA criteria -- negligible reflection and near-vanishing outside energy during excitation -- are clearly satisfied.

The incident wave arrives at the inside region  at $t\approx1000$, most of the transmission occurs until $t\approx1110$ and most of the reflection starts at $t\approx1095$, close to the time that transmission ends. This behavior is similar to that shown in Fig. \ref{fig:Virtual_Transmission_Double_Barrier}. Due to the non-zero energy of the tail and the dispersion in the right region, the energy outside the rightmost of the barrier is never exactly zero, but close to it, $\sim 7.5\%$. As we stated, the tail energy corresponds to $\sim 6.5\%$ of the total energy of the incident wave, explaining most of the remaining energy in the right region. The remaining $1\%$ energy can be attributed to the effect of the long tail of the potential on the propagation of the incident wave. Thus, the earlier reflection that can be seen in the red signal that extends to $x_0=125$ at $t=1190$ is mainly caused by the high dispersion which is related to high dimensionality of the problem. Moreover, one should be aware of the fact that VA is a process that is only possible with carefully constructed initial wave packets and we propose some of the ways for 4 different cases in curved geometries. There might be one-sided VA modes in high-dimensional Tangherlini geometry possessing higher quality factor which enables one to construct better one-sided VA simulations for black holes.
Finally, note that the half-width of the potential for BHs is of order $5M$, which corresponds also to the spatial scale of variation of the potential. Thus only waves with a wavelength larger than this are substantially reflected. But ``sluggish returns'' from the barrier, as in Fig.~\ref{fig:IDA11_Overplot} for the rectangular barrier, require that the wavelength be tuned to the potential height, $V_0 \sim 1/M^2$. We thus find that BHs are unable to display ``sluggish returns''.
\section{Discussion}\label{sec:discussion}
We have shown how dissipative systems can be perfect absorbers of radiation with a certain time-dependent profile. For stars, one can think about this unique process as the time-reversal of the characteristic relaxation of an object: instead of emitting radiation in certain characteristic modes, the system absorbs and stores the energy within it. BHs also have characteristic complex modes which can be excited to provide virtual total absorption. This is a new, rather unexplored feature of gravitational systems and time-dependent scattering.

The design of systems which can efficiently absorb incoming wavelike disturbances is of great value and importance in a range of technological applications, from radar detection to sound proofing, energy harvesting, etc.~\cite{https://doi.org/10.1002/adma.201200674,PhysRevApplied.3.037001,doi:10.1126/science.289.5485.1734,BARAVELLI20136562}. 
The use of virtual absorption in spherically symmetric 4 dimensional geometries is restricted. It is tempting, and amusing, to use VA to grow the energy content of a compact object to the point of collapse. One can think of illuminating a star with a long-lived QNM (hence a VA mode close to the real axis). Irradiating the star with a low-flux source over a very extended period of time, would presumably lead to virtual absorption to the point where nonlinearities become important. These could either then cause a total disruption of the star or collapse to a BH. The non-existence of compact objects with a radius below the light ring~\cite{Cardoso:2019rvt} indicates that a smooth adiabatic growth of the object to a BH is impossible.

Moreover, virtual total absorption phenomenon can be used for wave particles such as electrons in a double potential barrier system or a system with a Dirichlet boundary. In Quantum Mechanics, we can relate modes by a $E=\omega^2$ relation. Virtual absorption modes correspond to complex energies, exponentially decaying/growing sinusoidal wave function in time domain.
\begin{acknowledgments}
We are indebted to Valentin Boyanov and Elisa Maggio for useful discussions and for comparison of results concerning the modes of compact stars. We thank Emanuele Berti, Greg Cook, Saso Grozdanov, Hayato Motohashi, Naritaka Oshita, Paolo Pani, Sebastian Volkel and Mile Vrbica for interesting correspondence.
The Center of Gravity is a Center of Excellence funded by the Danish National Research Foundation under grant No. 184.
We acknowledge support by VILLUM Foundation (grant no. VIL37766) and the DNRF Chair program (grant no. DNRF162) by the Danish National Research Foundation.
F. T. is supported by TÜBİTAK under program 2224-d.
V.C.\ is a Villum Investigator and a DNRF Chair.  
V.C. acknowledges financial support provided under the European Union’s H2020 ERC Advanced Grant “Black holes: gravitational engines of discovery” grant agreement no. Gravitas–101052587. 
Views and opinions expressed are however those of the author only and do not necessarily reflect those of the European Union or the European Research Council. Neither the European Union nor the granting authority can be held responsible for them.
This project has received funding from the European Union's Horizon 2020 research and innovation programme under the Marie Sklodowska-Curie grant agreement No 101007855 and No 101131233.
This work is supported by Simons Foundation International \cite{sfi} and the Simons Foundation \cite{sf} through Simons Foundation grant SFI-MPS-BH-00012593-11.

\end{acknowledgments}

\appendix
\section{VA or TT modes of higher dimensional BHs}\label{app:larged}
Before demonstrating VA in spherically symmetric spacetimes, it is important to clarify which VAMs are available. In the existing literature, the only VAMs\footnote{Here we use the term VAM to refer to complex \textit{total transmission modes}.} known for four-dimensional spherically symmetric BHs are the so-called algebraically special modes in Schwarzschild spacetime~\cite{Couch:1973zc}. These modes are purely imaginary and, as a result, are not well suited for time-domain scattering simulations. We provide some details here of the calculation of the VA modes of higher dimensional BHs. Surprisingly, VA modes have only been calculated for Kerr BHs, and very recently only~\cite{Cook:2022kbb}. We suspect that there is a rich structure of modes in generic BH spacetimes. Here we focus on a very specific family of BH solutions in higher dimensions, the Tangherlini solution~\cite{Tangherlini:1963bw}.
\subsection{Perturbation theory on spherical symmetric BH spacetimes}
We consider a spherically symmetric spacetime in $d$ dimensions. In the usual Schwarzschild coordinates $x^\mu=(t,r,X^A)$ ($A=2\cdots d-1$), we work with the line element in the form
\beq
ds^2 =-f(r) \mathrm{d}t^2+\dfrac{1}{f(r)}\mathrm{d}r^2 +r^2 \mathrm{d}X^2,
\eeq
with $\mathrm{d}X^2$ the volume element of the unit sphere in $d-2$ dimension and
\be
f=1-\frac{r_h^{d-3}}{r^{d-3}}\,.
\ee
The horizon location in these coordinates is at $r=r_h$.
Black-hole perturbation theory formulated on the above spacetime leads to the radial master function~\cite{Kodama:2003jz,Kodama:2003kk}
\beq
\Bigg[ \dfrac{\mathrm{d}^2}{\mathrm{d}x^2}+\omega^2-V_s \Bigg] \psi(r) = 0.
\eeq
The effective potential depends on the spin $s$ of the field and on the angular number $\ell$ used to perform separation of angular variables~\cite{Kodama:2003jz,Kodama:2003kk,Cardoso:2003vt,Matyjasek:2021xfg},
\beq
V_s&=&f\biggl\{ \frac{\ell(\ell+d-3)}{r^2}+\frac{(d-2)(d-4)}{4r^2}\nonumber\\
&+&(1-s^2)\frac{(d-2)^2r_h^{d-3}}{4r^{d-1}}\biggr\} \,.
\label{Tangherlini_Potential}
\eeq
For $s=0$ the potential describes massless scalar fields, while for $s=2$ it describes a ``Regge-Wheeler''-like sector of gravitational perturbations. The tortoise coordinate is defined as usual via
\beq
\dfrac{\mathrm{d}x}{\mathrm{d}r} = \dfrac{1}{f(r)}.
\eeq

Virtual absorption modes (VAMs, as defined in this work, in BH literature they are also called total transmission) modes $\omega_\pm$ are defined according to the boundary conditions as $x\to \pm \infty$
\begin{eqnarray}
&\psi^- \sim e^{i\omega_- x}, \quad \rm{VA}\, (or \,TTM\, Left) \, \\
&\psi^+ \sim e^{-i\omega_+ x}, \quad  \rm{VA} \,(or\, TTM\, Right).
\end{eqnarray}
In other words, at TTM Left behaves as $\sim e^{i\omega_- x}$ at {\it both} boundaries.
\subsection{Compact coordinate adapted to VAMs}\label{sec:CompCoord_Numerics}
To solve the VAM problem using techniques similar to the conformal framework for BH theory~\cite{Zenginoglu:2011jz,PanossoMacedo:2023qzp}, we introduce a new set of coordinate system $x_\pm^\mu=(t_\pm, \sigma, \theta, \varphi)$ via
\beq
\label{eq:coord_trasfo}
t = r_h \bigg(t_\pm \mp x_d(\sigma)  \bigg), \quad r = \dfrac{r_h}{\sigma},
\eeq
with
\beq 
x_d(\sigma) = \dfrac{x(r(\sigma))}{r_h}
\eeq
the dimensionless tortoise coordinate. It is straightforward to see that $t_+=v/r_h$ and $t_-=u/r_h$, i.e., the new time coordinates are respectively the dimensionless ingoing and outgoing null coordinates. In the former case, $\sigma =0$ corresponds to past null infinity and $\sigma=1$ to the BH horizon, whereas in the latter $\sigma=0$ and $\sigma=1$ locate, respectively, future null infinity and the white hole horizon.

The time transformation \eqref{eq:coord_trasfo} implies the re-scaling of the frequency domain field~\cite{Zenginoglu:2011jz,PanossoMacedo:2023qzp}
\beq
\bar\psi^\pm(\sigma) = Z^\pm(\sigma)\,\psi^\pm(r(\sigma)), \quad Z^\pm(\sigma) = e^{\mp s_\pm x_d(\sigma)},
\eeq
with $s_\pm = -i \omega_\pm r_h$ a dimensionless re-scaling of the VAM frequency. The re-scaled field satisfies the generalised eigenvalue problem
\beq
\label{eq:QNM_EV}
L_1 [\bar \psi^\pm] = s_\pm L_2[\bar \psi^\pm]
\eeq
with
\beq
L_1 = \dfrac{d}{d\sigma}\left( p(\sigma) \dfrac{d}{d\sigma} \right) - \bar V_s(\sigma), \quad L_2 = 2\dfrac{d}{d\sigma}.
\eeq
The metric function $p(\sigma)$ and the conformal potential are defined by
\beq
p(\sigma) &=& -\dfrac{1}{x_d'(\sigma)} = \sigma^2 f(r(\sigma)), \\
\bar V_s(\sigma) &=& \dfrac{r_h^2}{p(\sigma)} V_{\ell m}(r(\sigma))
\eeq

Given the singular nature of $p(\sigma)$ at the boundaries $\sigma=0$ and $\sigma=1$, the VAM arises for the regular solutions $\bar \psi^\pm$ of eq.~\eqref{eq:QNM_EV}. We solve the eigenvalue problem numerically after discretizing the differentiation operators $L_1$ and $L_2$ via a spectral method based on Chebyshev collocation point~\cite{Jaramillo:2020tuu}. Since Left and Right modes relate via $s_- = - s_+$, we show results on the Left modes.

We complement the above procedure with a continued fraction implementation of the eigenvalue problem in the special five-dimensional case~\cite{Cardoso:2003vt,Matyjasek:2021xfg}. We find agreement to 10 decimal places, for the typical resolutions we used.

\subsection{Results}
We calculate the VAMs with the scheme described in sec.~\ref{sec:CompCoord_Numerics}. For that purpose, we perform a Chebyshev spectral discretization of the differentiation matrices with two resolutions $N_1=N$ and $N_2=N-5$. With the resulting set of eigenvalues associated with each resolution, we filter modes with relative difference within ${\rm TOL}<10^{-10}$. We perform tests with $N=75$, $N=100$, $N=150$ and $N=200$, with results for the highest resolution being displayed, unless stated otherwise. 

In the four-dimensional case ($d=4$), we recovered the expected values corresponding to the algebraically special modes
\begin{equation}
r_h \omega_{\pm} = \mp i\, \dfrac{(\ell -1)\ell(\ell+1)(\ell+2)}{6}   
\end{equation}
at the imaginary axis in the gravitational sector ($s=2$). On the other hand, the solver does not yield any VAM for scalar perturbations $s=0$. These results provide a strong benchmark for the numerical scheme laid out in sec.~\ref{sec:CompCoord_Numerics}.

Unlike the four-dimensional case, we find that both massless scalar fields and gravitational fluctuations have VAMs for $d > 4$. Most notably, some of these modes have complex frequencies. 
The next section summarizes the results for the gravitational and scalar sectors.

\subsubsection{spin 2}
%
For $s=2$, we find different families of gravitational VAMs. One with a purely imaginary component, akin to the algebraicaly special mode in $d=4$. Table~\ref{tab:spin2_lw} shows values for $\ell=2$ and $d\in[4,10]$.

\begin{table}[ht!]
    \centering
    \caption{Purely imaginary VAM for ${\rm spin}=2$ and $\ell =2$}
    \begin{tabular}{c|c}
        $d$ & $r_h \omega$  \\
        \hline
        \hline
        $4$ &  $4.00000000000 \, i$ \\
        $5$ &  $1.89632443230\, i$\\
        $6$ &  $1.50000000000\, i$ \\
        $7$ &  $1.34133678143\, i$\\
        $8$ &  $1.25747140951\, i$\\
        $9$ &  $1.20607423809\, i$\\
        $10$ & $1.17151998044\, i$ \\
    \end{tabular}
    \label{tab:spin2_lw}
\end{table}
At least some of these modes seem to have compelling numerical values (e.g. $d=6$), too compelling to be just a numerical coincidence. Given the absence of known ``superpartner relations'' as in the four-dimensional case, it is tempting to conclude that there is some yet unknown structure in these modes and possibly some hidden symmetries in the effective potential.

\begin{figure}[hb!]
\centering 
\includegraphics[width=0.95\columnwidth]{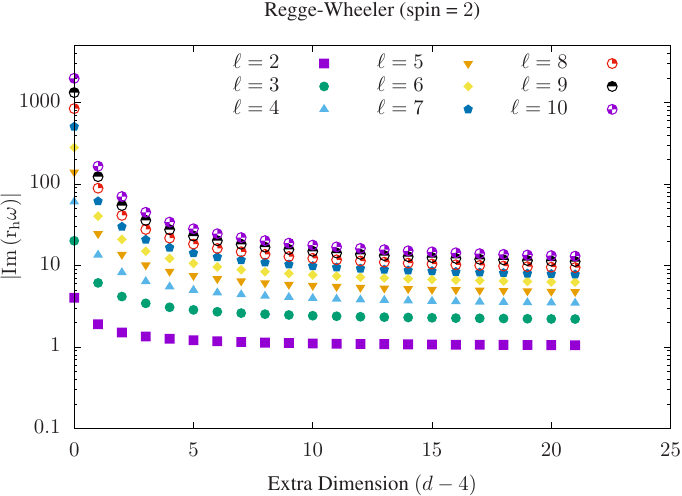} 
\includegraphics[width=0.95\columnwidth]{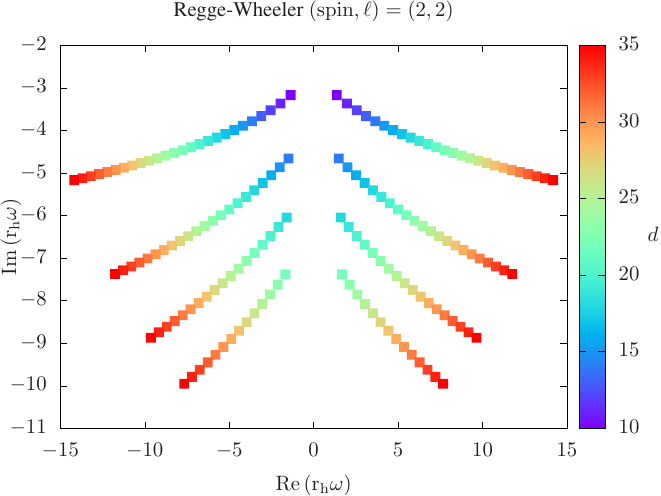}
 \caption{{\bf Top Panel:} Gravitational VAMs on the imaginary axis for the Tangherlini spacetime Regge-Wheeler potential (spin$=2$) for $\ell\in[2,10]$. For $d=4$ the values coincide with the well-known algebraically special modes. For $d\to \infty$ they seem to asymptote to a constant value. {\bf Bottom Panel:} Apart from the purely imaginary modes, we also find complex modes for $d\geq 10$. 
}
\label{fig:TTM_RW_Tangherlini_spin2}
\end{figure}

Besides, we also encounter a second family of modes for $d\geq 10$ with genuinely complex frequencies. Our results are summarized in Fig.~\ref{fig:TTM_RW_Tangherlini_spin2}. The top panel of the figure focuses on the purely imaginary modes for $\ell \in [2,10]$. We observe the VAMs asymptote a fixed value in the large $d$ regime. For instance, when $d\to \infty$, one finds approximately $\omega r_h \approx 1.048\,i$ for $\ell=2$.

The bottom panel displays the second family of gravitational VAMs with complex values when $d\geq 10$. As the dimension increases, new branches VAMs appear with larger $\left| {\rm Im} (r_h \omega) \right|$.

%
%

\subsubsection{spin 0}


The $s=0$ sector does not exhibt modes with purely imaginary values. However, we also find complex VA modes for the scalar field when $d\geq 10$. These are shown, for some values of spacetime dimension, in Table~\ref{tab:large_d}. 

\begin{table}[th!]
\caption{Comparison between numerically calculated value of scalar VA modes on a Tangherlini background and the large-d approximation. It is apparent that the large-$d$ approximation is increasingly accurate at large $d$. \label{tab:large_d}}
\addtolength{\tabcolsep}{-0.4em}
\begin{tabular}{c c c c}
\hline
$d$ & $\omega_{\mathrm{num}}$  &    $\omega_{\mathrm{ld}}$  & error  \\
\hline
60  & 27.348 + 6.196\,i    & 27.309 + 5.670\,i   & 1.9\%  \\
100 & 46.968 + 7.142\,i    & 46.842 + 6.653\,i   & 1.1\% \\
200 & 96.216 + 8.713\,i    & 96.020 + 8.251\,i   & 0.52\%\\
300 & 145.647 + 9.812\,i   & 145.426 + 9.358\,i  & 0.35\%  \\
400 & 195.182 + 10.685\,i  & 194.947 + 10.235\,i & 0.26\%\\
\hline
\end{tabular}
\end{table}

We can compare our numerics against a large $d$ approximation of the relevant equation~\cite{Emparan:2014cia}. We use a slightly more direct approach than Ref.~\cite{Emparan:2014cia}, by expressing the solution directly in terms of Hankel functions. At large $d$, the spin $0$ potential can be approximated by,
\be
V=\frac{\nu^{2}-\frac{1}{4}}{x^{2}}\Theta(x-x_{0})\,,
\label{Large_D_Potential}
\ee
where $\nu=\ell+\frac{d-3}{2}$ and $x_0\sim r_h$. We solve for the wave function,
\begin{equation}
\psi(r)=
\begin{cases}
e^{-i\omega (x-x_0)}, & x<x_0,\\[4pt]
\sqrt{x}\,\Big[ A_{\rm in}H_{\nu}^{(2)}(\omega x) + A_{\rm out}\,H_{\nu}^{(1)}(\omega x) \Big], & x>x_0.
\end{cases}
\end{equation}
where $H_{\nu}^{(1),(2)}$ are Hankel functions and $A_{\rm in}$, $A_{\rm out}$ correspond to incident and reflected amplitudes respectively.

We solve for $A_{\rm out}=0$ by imposing continuity for $\psi$ and $\psi'$ at $x_0\sim r_h$ to find VAMs and we compare the result with VAMs found by numerical solution with exact potential given by Eq.~\eqref{Tangherlini_Potential}. Our results are shown in Table \ref{tab:large_d}, showing remarkable agreement between numerical results and the analytical, large-$d$ approximation, lending further support to both approaches.
%

%
%
\bibliography{References}
\end{document}